\documentclass[twocolumn,superscriptaddress,showpacs, longbibliography, aps, prb]{revtex4-1}
\usepackage{color}
\usepackage{hyperref}
\usepackage{amsmath, amssymb}
\usepackage{ulem}
\hypersetup{colorlinks=true,breaklinks,linkcolor=blue,urlcolor=blue,citecolor=blue}
\usepackage{graphicx}% Include figure files
\usepackage{dcolumn}% Align table columns on decimal point
\usepackage{bm}% bold math
\usepackage{xspace}
\usepackage{physics}

\def\vec#1{\boldsymbol #1}

\newcommand{\sP}{\mathcal{P}}

\newcommand{\ra}{\rangle}
\newcommand{\Ns}{N_{\text{s}}}

\usepackage{natbib}

\begin{document}

\preprint{}

\title{
Correlated Zak insulator in organic antiferromagnets
}

\author{Takahiro Misawa}
\affiliation{Beijing Academy of Quantum Information Sciences, Haidian District, Beijing 100193, China}
\author{Makoto Naka}
\affiliation{School of Science and Engineering, Tokyo Denki University, {Ishizaka}, {Saitama 350-0394, Japan}}

\date{\today}

\begin{abstract}
Searching for topological insulators in solids is one of
the main issues of modern condensed matter physics
since robust gapless edge or surface states of the topological insulators 
can be used as building blocks of next-generation devices.
Enhancing spin-orbit couplings is a promising way to realize topological insulators in solids,
whereas the amplitude of the spin-orbit couplings is not sufficiently large 
in most materials. 
Here we show a way to realize a topological 
state characterized by the quantized Zak phase, termed the Zak insulator,
with spin-polarized edges in organic antiferromagnetic Mott insulators 
without relying on the spin-orbit coupling.
The obtained Zak insulator can have a large charge gap
compared to the conventional topological insulators,
since Coulomb interactions mainly govern 
the amplitude of the charge gap in the antiferromagnetic Mott insulators.
Besides the mean-field analysis, 
we demonstrate that the 
Zak insulator survives 
against electron correlation effects
by calculating the many-body Zak phase.
Our finding provides an unprecedented way to realize a topological 
state in strongly correlated electron systems. 
\end{abstract}

\maketitle
{\it Introduction.---}Topologically protected gapless edge or surface states, 
which universally appear in the topological insulators,
can propagate the charge or spin current with low dissipation~\cite{Hasan_RMP2010,Qi_RMP2011,Ando_JPSJ2013}.
A prototypical example of the topological insulators with edge states
is the quantum Hall systems under a high magnetic field~\cite{Klitzing_PRL1980}, 
where the topological invariant guarantees
the quantized Hall conductance and the gapless edge state~\cite{Thouless_PRL1982,Kohmoto_ANN1985,Hatsugai_PRB1993}.
The theoretical finding of another class of topological insulators, i.e., 
the $Z_{2}$ topological insulators 
in two~\cite{KaneMele} and three dimensions~\cite{FuKaneMele_PRL2007,Moore_PRB2007,Roy_PRB2009}, has stimulated searches for new types of topological insulators.
Nowadays, the periodic table of the topological insulators is established~\cite{Ryu_NPJ2010,Chiu_RMP2016}, 
which clarifies what kind of topological insulators can be realized 
in a given spatial dimension and symmetry of the system.
Efficient ways for identifying the topological insulators based 
on the symmetries of systems have also been 
developed and used to search for new topological materials~\cite{Kruthoff_PRX2017,Po_NCom2017,Bradlyn_Nature2017,Vergniory_Nature2019,Tang_NPhys2019,Tang_SA2019,
Zhang_Nature2019,Tang_NPhys2019,Xu_2020Nature,Iraola_PRB2021}.

A huge amount of work on topological insulators has been done
for mainly non-interacting (or weakly correlated) electron systems.
In the correlated electron systems, there are several proposals for the
topological states of the matter induced by the correlation effects.
One prominent example is the fractional quantum Hall system~\cite{Tsui_PRL1982},
where the electronic correlations induce the fractionalization
of the fermionic degrees of freedom.
Another example is the Kitaev spin liquid~\cite{Kitaev_ANP2006}, which may
realize in Mott insulators with strong spin-orbit couplings (SOC)~\cite{Jackeli_PRL2009}.
In the Kitaev spin liquid, the anisotropy in the magnetic interactions
induces the fractionalization of the
spin degrees of freedom, and it is shown that
the Majorana particles appears in the low-energy excitations.
Besides those mentioned above,
there are several other theoretical proposals of the
correlated topological phases, such as 
the topological Mott insulators~\cite{Raghu_PRL2008,Pesin_NPhys2010}
induced by the 
cooperation of SOC and electron correlation effects
and the magnetic Chern insulator~\cite{Martin_PRL2008}/Weyl semimetal~\cite{Shindou_PRL2001}
due to the non-coplanar magnetic orders.
Thus, most correlated topological insulators require 
strong SOC (or effective SOC induced by the electron correlations)
for their realization, as in the non-interacting topological insulators.

\begin{figure}[t] 
\begin{center} 
\includegraphics[width=0.5 \textwidth]{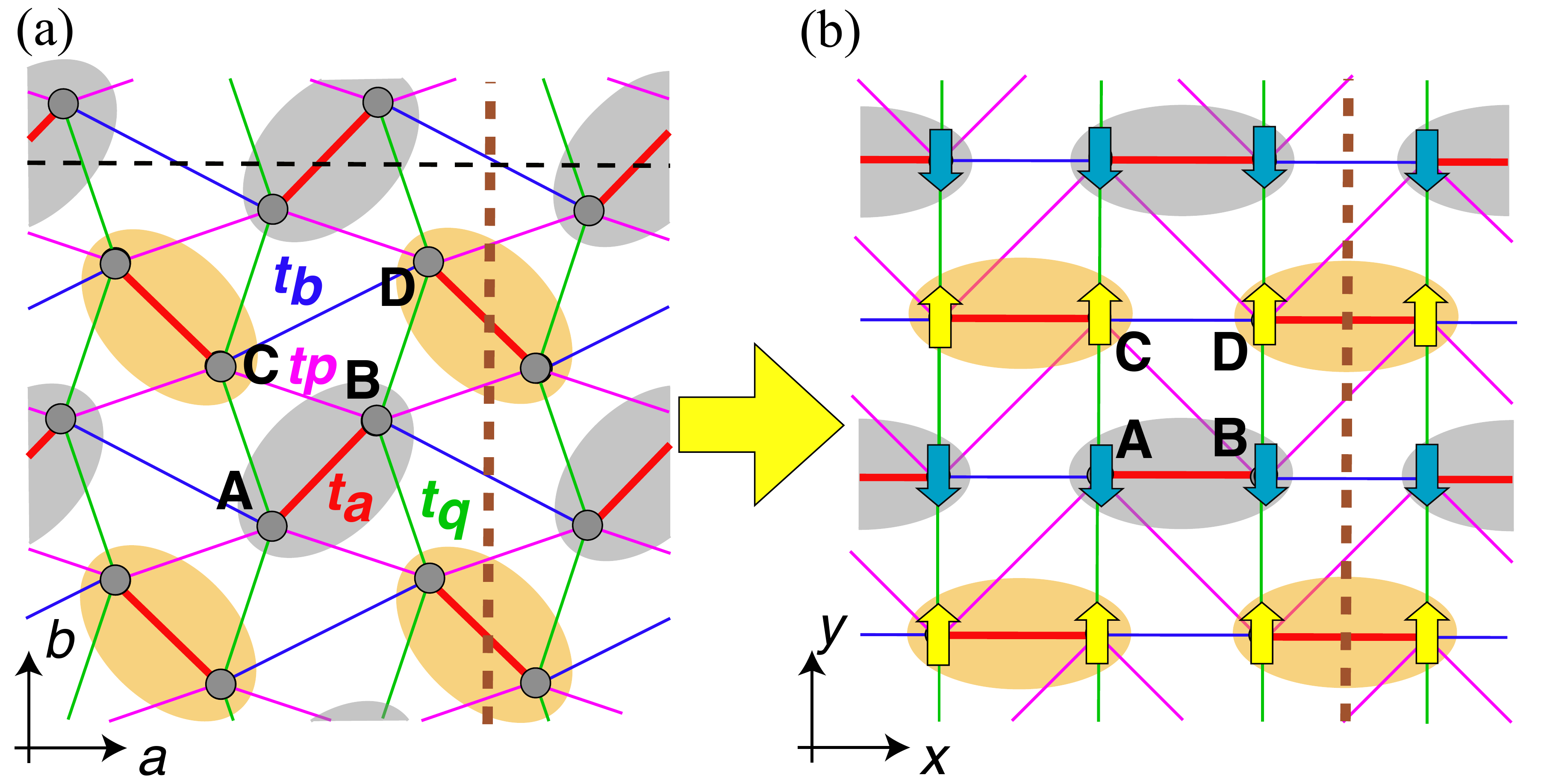}
\caption{(a)~Schematic lattice structures of {$\kappa$-$X$}. 
Shaded ovals represent the
BEDT-TTF dimers. A-D represents the 
independent BEDT-TTF molecules in the unit cell. 
The red, purple, green, and blue solid lines, denoted by $t_a$, $t_p$, $t_q$, and $t_b$, respectively, 
are the dominant intermolecular hopping integrals~\cite{Kino_JPSJ1996}.
The broken vertical (horizontal) line shows a truncation to generate
the gapless edge states parallel to the $b$ ($a$) axis 
(see Discussion in the main text).
(b)~Deformed lattice structure for clarifying the SSH chains emerge along the $x$ axis. 
The up and down arrows represent local spin moments in the AFM phase.
}
\label{fig:lattice}
\end{center}
\end{figure}

The topological insulators listed above are strong topological insulators, which
have gapless edge states regardless of the directions of truncations.
Other than the strong topological insulators, 
recent studies reveal the existence of weak topological insulators, 
which have gapless edge states in only certain directions of truncations.
A well-known example is the weak $Z_{2}$ topological insulator in
the Fu-Kane-Mele model~\cite{FuKaneMele_PRL2007}.

In this Letter, we demonstrate that an organic antiferromagnetic (AFM) Mott insulator
$\kappa$-(BEDT-TTF)$_{2}$Cu[N(CN)$_{2}$]$X$~\cite{Williams_IChem1990,Miyagawa_ChemRev2004,Kagawa_Nature2005,Kawasugi_NCom2016,Oike_PRL2015} 
(abbreviated as $\kappa$-$X$, and $X$ represents an anion taking Cl or Br) is
a correlated weak topological insulator 
{that is characterized by the quantized Zak phase, {termed the Zak insulator,}}
without relying on the SOC.
In the last few decades, organic conductors have been studied in terms of a 
{\it model compound} of 
strongly correlated electron systems, because of their simple electronic structures, 
compared to inorganic $d$ and $f$ electron systems. 
On the other hand, their topological properties have hardly been investigated, 
except for a small number of Dirac electron systems~\cite{Kitou_PRB2021,Ohki_PRB2022,Tanaka_PRL2022}, 
due to the weak SOC~\cite{Winter_PRB2017,Suzumura_JPSJ2021,Tsumuraya_EPJB2021}.
$\kappa$-{\it X} is one of the most well-studied Mott insulators~\cite{Kanoda_AR2011}, 
showing a wide variety of correlated phenomena, e.g., AFM orderings, superconductivity,  
and metal-to-insulator transitions. The crystal structure is composed of an 
alternate stacking of two-dimensional conducting BEDT-TTF layers and insulating anion {\it X} layers. 
Figure~\ref{fig:lattice}(a) shows the molecular arrangement in the conducting layer, 
where four BEDT-TTF molecules form two kinds of dimers with different orientations. 
The system has three electrons per dimer on average and three-quarter filled bands. 
In the following, we discuss that the combination of the dimer structure
and AFM ordering plays a key role in the emergence of a 
characteristic topological state.

First, we analyze the AFM insulating state in $\kappa$-$X$ by means of the mean-field approximation and
find that the spin-polarized gapless edge states appear 
at the edges that truncate the 
bonds with the strongest intradimer hopping integral $t_a$
and these
are characterized by the quantized Zak phase~\cite{Zak_RPL1989}.
Next, using the many-variable variational Monte Carlo method~\cite{Tahara_JPSJ2008,Misawa_CPC2019}, 
we calculate the many-body Zak phase described 
by the twist operator~\cite{Bohm_PR1949,Lieb_AP1961,Resta_PRL1998,Resta_PRL1999,Nakamura_PRL2002,
Watanabe_PRX2018,Balazs_PRR2020,Tasaki_arXiv2021} 
and confirm that the 
{Zak insulator} survives against quantum fluctuations 
and strong electron correlation effects.

\begin{figure}[t] 
\begin{center} 
\includegraphics[width=0.5 \textwidth]{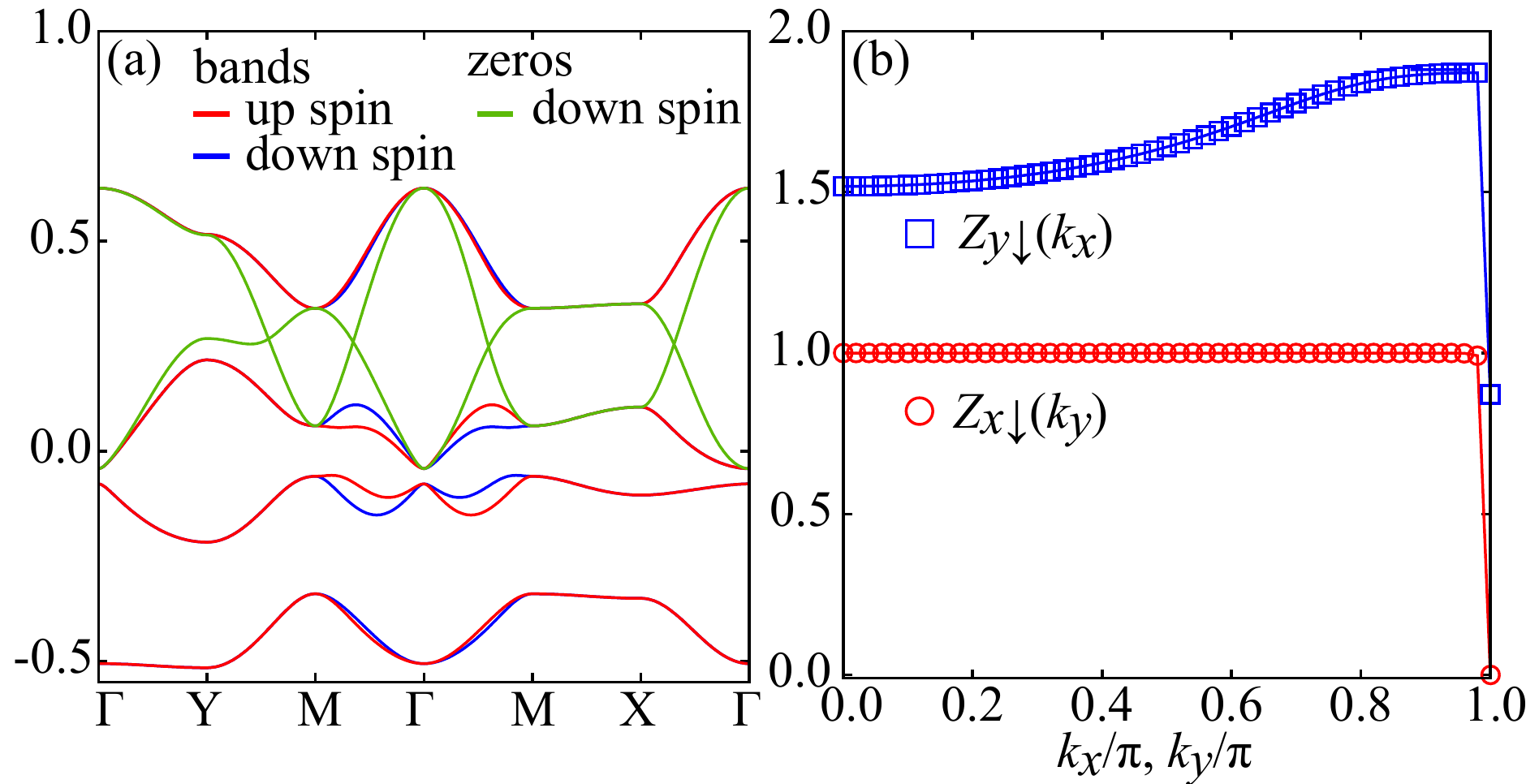}
\caption{(a)~Band structure 
of up-spin (red) and down-spin (blue) electrons in the AFM insulating state, 
obtained by the mean-field approximation for $\Delta=0.2$.
Green lines denote the zeros of the diagonal components of the Green functions for {down-spin} electrons.
For clarity, we only show the zeros
that traverse the band gap at three-quarter filling.
To see how the band inversion occurs, the unitary transformation
using the eigenvectors at $\Gamma$ point is also performed~\cite{Misawa_PRR2022}. 
(b)~$\bm k$ dependences of the Zak phase $Z_{\mu\downarrow}$ defined in Eq.~(\ref{eq:Zak}). 
We omit the real parts of $Z_{\mu\downarrow}$ because they are sufficiently small.
}
\label{fig:bulk}
\end{center}
\end{figure}

{\it Mean-field analysis.---}
In order to clarify how the edge state appears in the AFM state, 
we introduce the deformed lattice structure, 
where the A-D sites in Fig.~\ref{fig:lattice}(a) are mapped onto the square lattice,
as shown in Fig.~\ref{fig:lattice}(b).
In the deformed lattice, the intradimer ($t_{a}$) and 
interdimer ($t_{b}$) bonds align alternately along the $x$ axis.
The mean-field Hamiltonian in the AFM state, 
where the up-(down-)spin electrons locate on the C and D (A and B) sites as shown in Fig.~\ref{fig:lattice}(b), 
is given by
\begin{align}
&\mathcal{H}=\sum_{\vec{k},\sigma}\vec{c}^{\dagger}_{\vec{k}\sigma}H_{\sigma}(\vec{k})\vec{c}_{\vec{k}\sigma}, \\
&H_{\sigma}(\vec{k})=
\begin{pmatrix}
{\Delta}\sigma     ~~& R_{0}     ~~& R_{1}     ~~& R_{2} \\
R_{0}^{*} ~~& {\Delta}\sigma     ~~& R_{3}     ~~& R_{1} \\
R_{1}^{*} ~~& R_{3}^{*}   ~~& -{\Delta}\sigma  ~~& R_{4} \\
R_{2}^{*} ~~& R_{1}^{*}   ~~& R_{4}^{*} ~~& -{\Delta}\sigma
\end{pmatrix},
\label{eq:Ham}
\end{align}
where $\vec{c}_{\vec{k}\sigma}^{\dagger}=(c_{A\vec{k}\sigma}^{\dagger},c_{B\vec{k}\sigma}^{\dagger},c_{C\vec{k}\sigma}^{\dagger},c_{D\vec{k}\sigma}^{\dagger})$,
$R_{0}=t_{a}+t_{b}e^{-ik_{x}}$,
$R_{1}=t_{q}(1+e^{-ik_{y}})$,
$R_{2}=t_{p}(e^{-ik_{y}}+e^{-i(k_{x}+k_{y})})$,
$R_{3}=t_{p}(1+e^{ik_{x}})$,
$R_{4}=t_{b}+t_{a}e^{-ik_{x}}$,
and $\Delta$ is the gap induced by the AFM order.
The coefficient $\sigma$ takes $+1$ and $-1$ for up and down spins, respectively.
Following the $ab$ $initio$ evaluation~\cite{Koretsune_PRB2014}, 
we take the {intermolecular} hopping parameters for $\kappa$-Cl as
$(t_{a},t_{p},t_{q},t_{b})=(-0.207, -0.102, 0.043,-0.067)$ eV.
Figure~\ref{fig:bulk}(a) shows the band structure in the
AFM insulating {state} for {$\Delta=0.2$}, where the Fermi 
level is located in the gap at three-quarter filling.
{Since $\Delta$ is proportional to the amplitude of the onsite Coulomb interaction,
$\Delta$ can be sufficiently large to induce a large bulk gap
in AFM Mott insulators.}
We note that the spin splitting occurs in M-$\Gamma$ line 
due to the time-reversal symmetry breaking
even without the SOC, as is pointed out in the previous study~\cite{Naka_NCom2019}.

{\it Band inversion.---}
As discussed in Ref.~\onlinecite{Misawa_PRR2022},
the band inversion can be visualized by calculating the
zeros of the Green functions $G_{\sigma}(\omega,\vec{k})=(\omega I-H_{\sigma}(\vec{k}))^{-1}$,
where $\omega$ represents the energy.
We find that zeros of diagonal components of {the down-spin Green function} $G_{\downarrow}(\omega,\vec{k})$
traverses the insulating 
gap as shown in Fig.~\ref{fig:bulk}(a). 
This result indicates that the band inversion occurs
in the $k_{x}$ direction, i.e., 
between $\Gamma$ and $X=(\pi,0)$ or $M=(\pi,\pi)$ points.
The band inversion suggests that the down-spin bands in the $k_{x}$ direction have a topological nature.

{\it Quantization of the Zak phase.---}We examine the 
topological properties of the AFM insulating state 
by calculating the Zak phase~\cite{Zak_RPL1989} of the down-spin bands
both for the $k_{x}$ and $k_{y}$ directions, 
which is defined as
\begin{align}
{Z_{\mu\downarrow}}=-\frac{i}{\pi}\int_{-\pi}^{\pi} dk_{\mu}\Big[\langle u_{3\downarrow}({\vec{k}})|\frac{\partial}{\partial k_{\mu}}|u_{3\downarrow}({\vec{k}})\rangle\Big]
\label{eq:Zak}
\end{align}
where $|u_{3\downarrow}(k_{\mu})\rangle$ ($\mu=x,y$)
is the third eigenstate of $H_{\downarrow}(\vec{k})$.
We note that 
the Zak phase has been used
to identify the existence of the edge states 
in the analyses of the two-dimensional theoretical models~\cite{Ryu_PRL2002,Delplace_PRB2011}
and the $ab$ $initio$ calculations~\cite{Hirayama_NCom2017,Hirayama_PRX2018}.
Figure~\ref{fig:bulk}(b) shows
the Zak phase for the $k_{x}$ direction ({${Z}_{x\downarrow}$}), 
which is quantized to one except for $k_{y}=\pi$,
while ${Z}_{y\downarrow}$ is not.
These behaviors indicate the presence (absence) of the down-spin polarized edge states 
perpendicular to the $x$ ($y$) direction. 

\begin{figure}[t] 
\begin{center} 
\includegraphics[width=0.5 \textwidth]{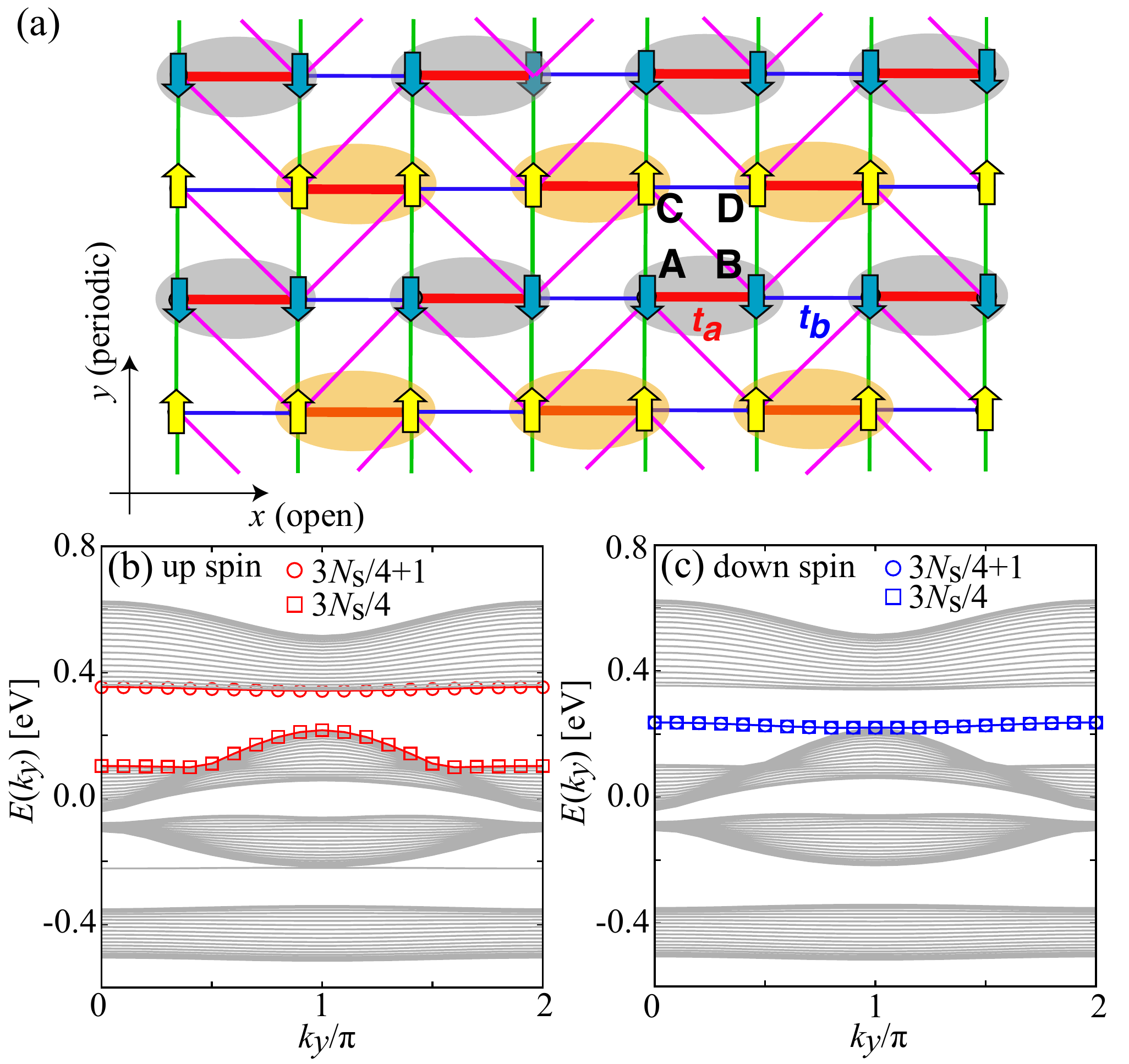}
\caption{(a)~
Schematic lattice structure for calculating the edge states. 
The open-(periodic-)boundary condition 
in the $x$ ($y$) direction is imposed.
Band 
{dispersions $E(k_{y})$} for (b) up- and (c) down-spin electrons. 
The blue squares and circles represent 
the gapless edge states of down-spin electrons 
at three-quarter filling.
}
\label{fig:edge}
\end{center}
\end{figure}

{\it Spin polarized edge state.---}We investigate the existence of the edge state
by calculating the band dispersions 
for the open boundary condition in
the $x$ direction
shown in Fig.~\ref{fig:edge}(a).
Figures~\ref{fig:edge}(b) and \ref{fig:edge}(c) show
the band dispersions for up- and down-spin electrons, respectively, 
where the gapless edge states appear only for the down-spin bands.
The down-spin edge states touch the bulk band dispersion at $k_y=\pi$.
This result is consistent with the non-quantized ${Z}_{x\downarrow}$ at $k_y=\pi$ in Fig.~\ref{fig:bulk}(b).

Here, 
we discuss the origin of the spin-polarized edge states in the AFM state.
For simplicity, we consider the strong coupling limit where the AFM
ordered moment is saturated.
Let us focus on the up-spin polarized one-dimensional chain along the $x$ axis 
composed of the C and D molecules. 
In the C-D chain, each dimer is occupied by the 
two up-spin electrons and one down-spin electron in the AFM state since the 
number of electrons per dimer is three; 
the up-spin states are fully occupied and the down-spin states are half-filled. 
By considering only the down-spin electrons, the C-D chain can be identified as the 
half-filled {\it spinless fermion system} on the dimerized one-dimensional chain, i.e., 
the Su-Schrieffer-Heeger (SSH) model~\cite{Su_PRL1979}, 
disregarding the interchain hoppings $t_p$ and $t_q$. 

In the lattice structure shown in Fig.~3(a), each C-D chain is truncated at the intradimer bond with $t_a$, 
which is equivalent to the topological state of the SSH model, 
where two unpaired molecules appear at both ends of the chain. On the other hand, 
the down-spin polarized A-B chains, which is regarded as the SSH model of the up-spin electron, do not have such unpaired molecules. 
As a result, only the down-spin polarized edge states appear.
The edge states of each chain are not independent of each other due to $t_p$ and $t_q$, 
which results in the band dispersion along the $k_y$ axis as presented in Fig.~3(c).
In the same manner, when the boundary condition 
truncating the $t_a$ ($t_b$) bonds 
in the A-B (C-D) chains is chosen, 
the spin polarization of the edge states is reversed.
This reversal occurs when the electron density is changed from 
three-quarter 
to quarter filling, 
where up-spin polarized
edge states appear 
around $-0.2$ eV between the two bonding bands as shown
in Fig.~3(b).

{\it mVMC analysis.---}To examine the correlation effects on the topological state
beyond the mean-field level, by using the mVMC method~\cite{Tahara_JPSJ2008,Misawa_CPC2019}, 
we analyze the following Hubbard-type
effective model for $\kappa$-$X$~\cite{Kino_JPSJ1996,Seo_JPSJ2000,Hotta_PRB2010,Naka_JPSJ2010,Seo_JPSJ2021}.
\begin{align}
&H=\sum_{ij,\sigma}t_{ij}(c_{i\sigma}^{\dagger}c_{j\sigma}+{\rm H.c.})
+U\sum_{i}n_{i\uparrow}n_{i\downarrow}\notag\\ 
&+V\sum_{i,j\in{\rm dimer}}n_{i}n_{j},
\label{eq:HubbardHam}
\end{align}
where $n_{i}=n_{i\uparrow}+n_{i\downarrow}$, 
$t_{ij}$ is the hopping parameters, and $U$ ($V$) is
the intramolecular (intradimer) Coulomb interaction.
For the intermolecular hopping parameters, 
we adopt the same values 
as the mean-field Hamiltonian 
in Eq.~(\ref{eq:Ham}).
The intradimer Coulomb interaction is set to $V=U/10$ as 
a typical value. 
We use the deformed lattice structure with $\Ns=(2L_{x})\times(2L_{y})$
both for the $x$ and $y$ directions, 
where $2L_\mu$ ($\mu=x,y$) denotes the number of molecules along the $\mu$ axis
under the periodic boundary condition.
The form of the variational wave functions 
is given by
\begin{align}
|\psi\ra =\sP_{\rm G}\sP_{\rm J}|\phi_{\rm pair}\ra,
\label{mvmc_wavefunction_def}
\end{align}
where $\sP_{\rm G}$ and $\sP_{\rm J}$
represents the Gutzwiller~\cite{Gutzwiller_PRL1963} and 
the long-range Jastrow factors~\cite{Jastrow_PR1955,Capello_PRL2005}, {respectively}.
The pair product wave function $|\phi_{\rm pair}\ra$  is defined as
\begin{align}
|\phi_{\rm pair}\ra= \Big[\sum_{i,j=0}^{\Ns-1}
f_{ij}{c_{i\uparrow}^{\dag}c_{j\downarrow}^{\dag}}\Big]^{N_{e}/2} |0 \ra,
\end{align}
where $f_{ij}$ represents the variational parameters and
$N_{\rm e}$ is the number of electrons.
We impose $2\times2$ sublattice structure in the variational parameters
to take into account the AFM order.
We optimize all the variational parameters simultaneously 
using the stochastic reconfiguration method~\cite{Sorella_PRB2001}. We use the particle-hole transformation
to reduce numerical costs in the actual calculations.

\begin{figure}[t] 
\begin{center} 
\includegraphics[width=0.5 \textwidth]{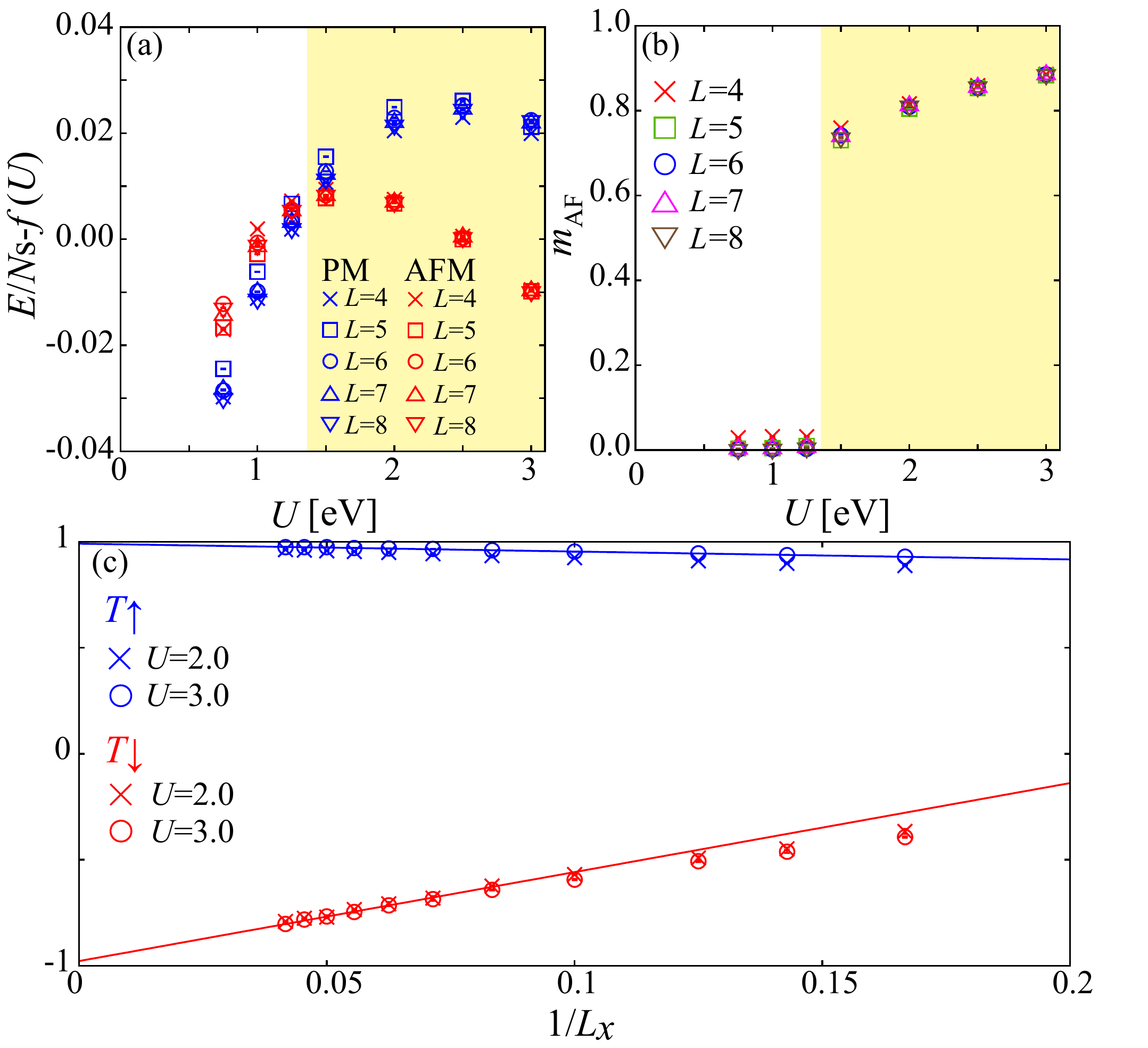}
\caption{(a)~Energies of the PM and AFM solutions
as a function of $U$
in the unit of eV, where the contributions proportional to $U$ are subtracted for clarity.
The system size is chosen as $L=L_{x}=L_{y}$.
(b)~The AFM order parameter $m_{\rm AF}$ as a function of $U$.
(c)~The expectation value of the twist operator defined
in Eq.~(\ref{eq:twist}). 
We fix $L_{y}=4$ and change $L_{x}$ from 6 to 24.
The lines indicate the results of the least square fitting of the data for $L_{x}\geq 18$ and $U=3$.} 
\label{fig:mvmc}
\end{center}
\end{figure}

We first examine the stability of
the AFM order in the Hamiltonian
defined in Eq.~(\ref{eq:HubbardHam}). As the initial states,
we choose an AFM state and a paramagnetic (PM) state.
By optimizing them, we evaluate 
energies of the
AFM and the PM states.
Figure~\ref{fig:mvmc}(a) shows
these energies as functions of $U$, where
the AFM state 
becomes the ground state for $U\gtrsim1.5$ {eV}.
The energy-level crossing indicates the first-order phase transition between 
the AFM and PM phases.
We also show the {AFM ordered} moment defined by
\begin{align}
&m_{\rm AF}=|S_{A}^{z}+S_{B}^{z}-S_{C}^{z}-S_{D}^{z}|,\\
&S_{\nu}^{z}=\frac{1}{L_{x}L_{y}}\sum_{i\in \nu}\frac{1}{2}(\ev*{n_{i\uparrow}}-\ev*{n_{i\downarrow}}),
\end{align}
in Fig.~\ref{fig:mvmc}(b), where
$m_{\rm AF}$ becomes finite above $U\sim 1.5$ {eV}.
Thus, this result shows that the ground state of the Hubbard model in Eq.~(\ref{eq:HubbardHam}) is
the AFM ordered state 
even if we take into account quantum and spatial correlation effects seriously 
beyond the mean-field approximation.

{\it Many-body Zak phase.---}Here, we examine the topological nature of the
AFM state obtained by 
the mVMC method. To extract the many-body topological invariant, 
we calculate the expectation value of the following twist operator, defined as
\begin{align}
\hat{T}_{\sigma}(L_{x},L_{y})=\exp[\sum_{j=0}^{N_{\rm s}-1}i\frac{2\pi}{L_{x}}x(j){{n}_{j\sigma}}],
\end{align}
where $x(j)$ is the $x$ coordinate of the unit cell, to which the $j$th site belongs.
This twist operator is {a} two-dimensional extension of the
one-dimensional twist operator for spinless fermions~\cite{Resta_PRL1998,Resta_PRL1999,Tasaki_arXiv2021} defined as
\begin{align}
\hat{T}_{\rm 1D}(L_{x})=\exp[\sum_{x=0}^{L_{x}-1}i\frac{2\pi}{L_{x}}x{{n}_{x}}].
\end{align}
For the one-dimensional SSH model,
the expectation value of $\hat{T}_{\rm 1D}$ in large system sizes becomes~\cite{Tasaki_arXiv2021}
\begin{align}
\ev*{\hat{T}_{\rm 1D}(L_{x})}=(-1)^{L_{x}+1}e^{i\nu\pi},
\end{align}
where $\nu$ is the Zak phase, which 
becomes one (zero) for the topological (trivial) phase.
Since $\hat{T}_{\sigma}$ can be regarded as the stacking of $\hat{T}_{\rm 1D}$,
in large system sizes, we obtain the following relation
\begin{align}
\ev*{\hat{T}_{\sigma}(L_{x},L_{y})}=\ev*{\hat{T}_{\rm 1D}(L_{x})}^{L_{\rm y}} = e^{i\nu\pi L_{y}}(-1)^{L_{y}(1+L_{x})}.
\end{align}
Thus, in the topological and trivial {phases},
we obtain $\ev*{\hat{T}_{\sigma}}=(-1)^{L_y(2+L_{x})}$
and $\ev*{\hat{T}_{\sigma}}=(-1)^{L_y(1+L_{x})}$, respectively.
Therefore, the topological (trivial) phase is characterized 
by $\langle \hat{T}_{\sigma} \rangle = -1$ ($\langle \hat{T}_{\sigma} \rangle = 1$)
when both $L_{x}$ and $L_{y}$ are odd.
Using the wave function obtained by 
the mVMC method, we evaluate the following quantity defined as
\begin{align}
T_{\sigma}(L_{x},L_{y})=\mathcal{S}_{\sigma}|\ev*{\hat{T}_{\sigma}(L_{x},L_{y})}|^{\frac{1}{L_{y}}},
\label{eq:twist}
\end{align}
where $\mathcal{S}_{\sigma}$ is a sign of $\ev{T_{\sigma}(L_{x},L_{y})}$
for odd $L_{x}$ and $L_{y}$, e.g.,
$\mathcal{S}_{\sigma}={\rm sign}(\ev*{T_{\sigma}(L_{x}=5,L_{y}=5)})$.
When the {electron with} $\sigma$ spin has the topological nature,
$T_{\sigma}(L_{x},L_{y})=-1$ is satisfied in large system sizes.
We note that thin-torus geometry is necessary to
correctly take the thermodynamic limit of 
the expectation values of the twist operators
in two or higher dimensions~\cite{Nakagawa_PRB2017,Watanabe_PRX2018}.
Thus, in actual calculations, we fix $L_y$ and change $L_x$
to take the thermodynamic limit.

In Fig.~\ref{fig:mvmc}(c),
we show size {dependences} of $T_{\sigma}(L_{x},L_{y})$
for $U=2$ and $3$, where the AFM
{state} is the ground state.
We find that $T_{\downarrow}$ is converged to $-1$ while $T_{\uparrow}$ is converged to $1$ in
the thermodynamic limit for both values of $U$. 
This result demonstrates that the down (up) spin electrons
have a topologically non-trivial (trivial) nature in the AFM state. 
The existence of the non-trivial many-body Zak phase 
only for down-spin electrons indicates that the spin-polarized edge states appear
in the AFM state, which is consistent with the results of the mean-field analysis.

{\it Discussion.---}In the 
{analysis} so far, we have considered the deformed lattice in Fig.~\ref{fig:lattice}(b) 
and focused on the edge states perpendicular 
to the $x$ axis,~i.e., $a$ axis in the original lattice in Fig.~\ref{fig:lattice}(a),
for simplicity. 
On the other hand, in the original lattice, 
since the A-B and C-D dimers are oriented by almost $45$ degrees as shown in Fig.~\ref{fig:lattice}(a), 
we can choose the edge perpendicular to the 
$b$ axis to 
truncate the strong $t_{a}$ bonds
under the open (periodic) boundary condition for the 
$b$ ($a$) direction, as exemplified by
the thin (black) broken line in Fig.~\ref{fig:lattice}(a).
In this case, the spin-polarized edges states emerge 
perpendicular to the $b$ axis.

These spin-polarized edge states originate from the cooperative effects of the dimer structure and AFM ordering. 
Focusing on the up(down)-spin polarized dimers in the AFM ordered state, 
a half-filled fermion system with the up(down) spins in the 1D dimer chain, i.e., the SSH model, 
is spontaneously formed. In order to experimentally realize a half-filled spinless fermion system, 
one requires to spin-polarize a quarter (or three-quarter) filled electron system 
by applying a strong magnetic field which sometimes leads to additional changes in the electronic state.
On the other hand, the present result shows that it can be more easily realized 
owing to electron correlation effects without any external field. 
The dimer structure, 
three-quarter-filled band, and the AFM ordering are widely common not only in $\kappa$-type ET salt 
but also in many other organic charge-transfer complexes, e.g., TMTTF, dmit, and $\beta^{\prime}$-type ET salts. 
These systems will provide useful platforms for future experimental studies on the 
correlated Zak insulator with the spin-polarized edge states.

We emphasize that the present mechanism is strikingly 
different from the so-called topological Mott insulator in Ref.~\onlinecite{Raghu_PRL2008}, 
where the complex bond order parameter, via the Fock decoupling of 
the intersite Coulomb interaction, activates the topological phase. 
In this study, $V$ plays a supplementary role in stabilizing the AFM Mott insulating state 
and is not essential for the emergence of the topological state;~as far as the 
AFM insulating state remains, 
the topological insulator with the quantized Zak phase 
can realize without $V$.

We comment on the 
classification of the topological phase found in
this study. In this Letter, we use the term `topological' insulator
in the sense that `insulator with a non-trivial topological invariant'.
In this sense, 
the AFM Mott insulator in $\kappa$-$X$ can be regarded as a correlated `topological' insulator
with the quantized Zak phase. This is the reason
why we call the AFM Mott insulator the correlated Zak insulator.
However, 
the quantization of the Zak phase relies on the particular surface termination, 
truncating the intra-dimer bond.
Thus, strictly speaking, this Zak insulator is not a strong topological insulator 
but classified into a weak one. 
Besides truncating the intra-dimer bond, the edge state might be achieved 
by chemical substitution of one ET molecule of the dimer on the surface.

Here we discuss experimental detections of the present spin-polarized edge states.
The experimental techniques sensitive to surface magnetization and 
applicable to organic compounds are considered to be helpful, e.g., 
magneto-optical Kerr effect~\cite{Kato_2004Science} and magnetic force microscopy~\cite{Kazakova_JAP2019}.
Whether the net surface magnetization survives or not strongly depends 
on the interlayer stacking of the two-dimensional AFM orders 
along the $c$ axis, which is perpendicular to the $a$ and $b$ axes.
According to the recent experiments\cite{Ishikawa_JPSJ2018,Oinuma_PRB2020},
there are two AFM compounds, $\kappa$-Cl and 
the deuterated $\kappa$-Br, showing different stacking patterns.
Applying the present results to these compounds, we can expect that $\kappa$-Cl shows 
the net surface magnetization {along the $a$ axis, which is perpendicular to the bulk weak ferromagnetic 
moment along the $b$ axis due to the DM interaction,} while $\kappa$-Br does not, 
depending on their AFM stacking patterns.
Therefore, the comparison between these two compounds 
provides a good testbed for our scenario in experiments.

Finally,  
we refer to the possibility of the reconstruction
of the edge states, which might occur in the strongly correlated region but is not considered in our study.
To accurately investigate how the reconstruction occurs, it is necessary to
examine the effects of the long-range part of the Coulomb interactions in the effective Hamiltonians.
The $ab$ $initio$ downfolding method~\cite{Aryasetiawan_PRB2004,Imada_JPSJ2010,Nakamura_CPC2021}
is a promising way to accurately evaluate the screened Coulomb interactions in the low-energy effective Hamiltonians.
In previous studies, it has been shown that the $ab$ $initio$ effective Hamiltonians 
can correctly describe {the} electronic structures of several molecular solids~\cite{Shinaoka_JPSJ2012,Misawa_PRR2020,Yoshimi_PRR2021,Ido_npj2022}.
Performing such an $ab$ $initio$ investigation for $\kappa$-$X$ 
is an intriguing issue but left for future studies.

\begin{acknowledgments}
TM wishes to thank Youhei Yamaji for fruitful discussions 
and Kota Ido for providing a code for calculating twist operators.
{We also thak Tetsuya Furukawa and Takahiko Sasaki for fruitful discussions 
on experimental realization of the edge states.}
TM was supported by Building of Consortia for 
the Development of Human Resources in Science and Technology, MEXT, Japan.
This work was supported by a Grant-in-Aid for Scientific Research
No.~JP19K03723 from the Ministry of Education, Culture, Sports, Science and Technology, Japan,
and the GIMRT Program of the Institute for Materials Research, Tohoku University, No. 202112-RDKGE-0019. 
This work was also supported by the National Natural Science Foundation of China (Grant No. 12150610462).
\end{acknowledgments}
%\bibliography{Topo}

\providecommand{\noopsort}[1]{}\providecommand{\singleletter}[1]{#1}%

\end{document}